\documentclass[10pt]{iopart}

\usepackage{iopams}
\usepackage[dvips]{graphicx}
\usepackage{amssymb}

\newcommand{\rthz}[2]{#1~#2~$\mathrm{Hz}^{-1/2}$}

\begin{document}

\title[]{Measuring the LISA test mass magnetic proprieties with a torsion pendulum}

\author{M Hueller$^a$, M Armano$^{a,b}$, L Carbone$^a$, A Cavalleri$^c$,\\ R Dolesi$^a$, C D Hoyle$^a$\footnote[1]
{Currently at the Department of Physics, University of Washington,
Seattle, WA 98195 USA}, S Vitale$^a$ and W J Weber$^a$}

\address{$^a$ Dipartimento di Fisica, Universit\`a di Trento and INFN, Sezione di Padova, Gruppo Collegato
di Trento, I-38050, Povo, Trento, Italy}
\address{$^b$ Dipartimento di Fisica, Universit\`a dell'Insubria, I-22100, Valleggio, Como, Italy}
\address{$^c$ CEFSA-ITC, I-38050, Povo, Trento, Italy}

\ead{hueller@science.unitn.it}

\begin{abstract}Achieving the low frequency LISA sensitivity
requires that the test masses acting as the interferometer end
mirrors are free-falling with an unprecedented small degree of
deviation. Magnetic disturbances, originating in the interaction
of the test mass with the environmental magnetic field, can
significantly deteriorate the LISA performance and can be
parameterized through the test mass remnant dipole moment
$\vec{m}_r$ and the magnetic susceptibility $\chi$. While the LISA
test flight precursor LTP will investigate these effects during
the preliminary phases of the mission, the very stringent
requirements on the test mass magnetic cleanliness make
ground-based characterization of its magnetic proprieties
paramount. We propose a torsion pendulum technique to accurately
measure on ground the magnetic proprieties of the LISA/LTP test
masses.

\end{abstract}


\section{Introduction}
\label{introduction} The LISA (Laser Interferometer Space Antenna)
sensitivity goal requires that the test masses (nominally 2 kg)
are kept in free fall with an acceleration noise below \rthz{3}{fm
s$^{-2}$} at frequencies down to 0.1 mHz \cite{bender}. In order
to achieve this high purity of geodesic motion, environmental
noisy forces are screened by shielding the test masses in a
drag-free satellite, with precision thrusters driven by a position
sensor in order to minimize the test mass - satellite relative
displacement. Among the residual disturbance sources, magnetic
effects play a paramount role, as discussed in
\cite{LTPA:final,hueller:thesis,hanson:mag}: the fluctuations of
both magnetic field and magnetic field gradient will couple to the
test mass remnant dipole moment $\vec{m}_r$ and susceptibility
$\chi$, to produce force noise. In the limit of weakly magnetic
materials, the component of the force acting on the mass along the
LISA interferometer axis $x$ can be expressed as
\begin{equation}\label{e:mag:force:2}
f_{x}(t) \approx \left(\vec{m}_r + \frac{\chi V}{\mu_0}
\vec{B}(t)\right) \cdot
    \frac{\partial}{\partial x}
    \vec{B}(t) =
    \vec{m}_r \cdot \frac{\partial}{\partial x} \vec{B}(t)
    + \frac{\chi V}{\mu_0} \vec{B}(t) \cdot \frac{\partial}{\partial x} \vec{B}(t)
    \end{equation}with similar relations holding for $f_{y}(t)$ and $f_{z}(t)$.
Here we describe the test mass magnetic proprieties by its
permanent, remnant magnetic dipole moment $\vec{m}_r$ and its
magnetization induced through the (small) magnetic susceptibility
$\chi$ by the externally applied magnetic field $\vec{B}(t)$. For
LISA, fluctuations of $\vec{B}(t)$ are expected to be dominated by
the interplanetary magnetic field, while fluctuations of the
magnetic field gradient $\partial \vec{B}(t)/\partial x$ are
expected to be produced by sources on the satellite itself. In
order to relax the consequent environmental requirements on the
satellite, it is crucial to obtain a test mass with very good
magnetic proprieties; for LISA the requirements are $\left|
\vec{m}_r \right| \leqslant 10$~nA m$^2$ and $\left| \chi \right|
\leqslant 3 \cdot 10^{-6}$. The test mass design calls for a
70\%Au-30\%Pt alloy, with composition chosen in order to achieve
the lowest susceptibility, while retaining high density to
minimize the displacement caused by a given force disturbance. The
characterization of the full-sized test mass for LISA and its
flight precursor LTP~\cite{LTP:2}, particularly important given
the very stringent requirements on the magnetic cleanliness level,
is made difficult with the standard magnetic characterization
techniques, such as SQUID magnetometers and susceptometers, by its
relatively large dimensions.

In this article we discuss an application of a high sensitivity
torsion pendulum facility, developed for several testing of force
disturbances for
LISA~\cite{carbone:cqg,carbone:prl,carbone:cqg:2}, to the
independent characterization of both the test mass remnant moment
and susceptibility, assessing these proprieties directly through
the forces and torques associated with the variation of magnetic
fields.

\section{Characterizing a LISA test mass with a torsion
pendulum} \label{s:tp}In the proposed experiment we will measure
with high resolution the AuPt test mass remnant moment $\vec{m}_r$
and susceptibility $\chi$, by exploiting the high torque
sensitivity of a torsion pendulum where the test mass is included
in a light, non magnetic holder and is suspended by a thin fiber,
as sketched in \fref{f:mag:tor:pend:1}. The magnetic proprieties
of the LISA test mass will be measured by observing the torques
acting on it when subjected to a controlled oscillating field
$\vec{B}_{\textrm{f}}(\vec{x},t) = \vec{B}_0(\vec{x})
\sin{\omega_m t}$ produced by a suitable excitation coils
configuration. Assuming the addition of the external, residual dc
magnetic field $\vec{B}_{\textrm{res}}$, the total applied field
is then $\vec{B}(\vec{x},t) = \vec{B}_{\textrm{res}}(\vec{x}) +
\vec{B}_0(\vec{x}) \sin{\omega_m t}$. In order to evaluate the
effect of the applied magnetic field we need to account 3
components of $\vec{B}_0$ and $\vec{B}_{\textrm{res}}$, which
depend on position within the test mass $\vec{x} = (x_i,y_i,z_i)$.
The test mass is then modelled by meshing it into a grid of small
elements with volume $\mathcal{V}$, located at the positions
$\vec{x}$, each small enough to assume locally uniform field
$\vec{B}(t) =
(B_{x}(t,\vec{x}),B_{y}(t,\vec{x}),B_{z}(t,\vec{x}))$. Each
element $i$ interacting with the externally applied magnetic field
is characterized by a remnant magnetic moment $d
\vec{m}_r(\vec{x})$ (with $\sum_i d \vec{m}_r(\vec{x}) =
\vec{m}_r$) and a susceptibility $\chi(\vec{x})$. The force along
the $x$ axis acting on each test mass element can then be
expressed using \eref{e:mag:force:2} as a combination of a dc term
\begin{equation}\label{e:mag:force:dc}\nonumber
    {f_{x}}_{\textrm{dc}} \approx \left(d \vec{m}_r +
    \frac{\chi \mathcal{V}}{\mu_0}\vec{B}_{\textrm{res}}\right) \cdot
    \frac{\partial}{\partial x} \vec{B}_{\textrm{res}} +
    \frac{\chi \mathcal{V}}{2 \mu_0} \vec{B}_{0} \cdot \frac{\partial}{\partial x} \vec{B}_{0},
\end{equation}a term at the modulation frequency $f_m = \omega_m / 2 \pi$
\begin{equation}\label{e:mag:force:1w}\nonumber
    {f_{x}}_{1 \omega_m} \approx
    \left[\left(d \vec{m}_r + \frac{\chi \mathcal{V}}{\mu_0} \vec{B}_{\textrm{res}} \right)
    \cdot \frac{\partial}{\partial x}
    \vec{B}_{0} + \frac{\chi \mathcal{V}}{\mu_0} \vec{B}_{0} \cdot \frac{\partial}{\partial x} \vec{B}_{\textrm{res}}
    \right] \sin{\omega_m t}
    \end{equation} and a term at twice the modulation frequency $2 f_m$
\begin{equation}\label{e:mag:force:2w}
        {f_{x}}_{2 \omega_m} \approx -\frac{\chi \mathcal{V}}{2 \mu_0}\left(\vec{B}_{0}
    \cdot \frac{\partial}{\partial x} \vec{B}_{0}\right)\cos{2 \omega_m t}.
\end{equation}The torque around the vertical $z$ axis running
through the test mass center of mass is given instead by the
interaction of the horizontal ($x$,$y$) projections of the remnant
magnetic moment, $d \vec{m}_r$, with the total applied magnetic
field:\begin{equation}\label{e:mag:torque:1}
    \vec{n_z}(t) \approx (d \vec{m}_r \times
    \vec{B}(t))_z,
\end{equation}
and can be written as a combination of a dc term
\begin{equation}\label{e:mag:torque:dc}{\vec{n_z}}_{\textrm{dc}} =
(d \vec{m}_r \times \vec{B}_{\textrm{res}})_z
\end{equation} and a $1 f_m$ term
\begin{equation}\label{e:mag:torque:1w}{\vec{n_z}_{1 \omega_m}}(t) =
(d \vec{m}_r \times \vec{B}_{0})_z \sin{\omega_m t},
\end{equation}while the $2 f_m$ component vanishes because the
magnetic moment induced through the susceptibility $\chi$ is
parallel to the applied field. After evaluating the forces acting
on each test mass element $f_{x}(x_i,y_i,z_i)$ and
$f_{y}(x_i,y_i,z_i)$ given by \eref{e:mag:force:2}, together with
the torque about the $z$ axis $n_z(x_i,y_i,z_i)$ given by
\eref{e:mag:torque:1}, we account also for the torque arising from
the net force difference at different locations within the test
mass due to the remnant and induced magnetic moment:
\begin{equation}\label{e:mag:dx:force}
    \tilde{N_i}_{z}(t) \approx x_i f_{y}(t,x_i,y_i,z_i) - y_i f_{x}(t,x_i,y_i,z_i).
\end{equation}Here $x_i$ and $y_i$ are the horizontal distances of the
element from the torsional axis.

For given external and applied fields, the overall torque about
the $z$ axis $N_z$, which the torsion pendulum is sensitive to, is
then evaluated by adding the contributions from all the mass
elements $i$:
\begin{equation}\label{e:mag:total:torque} N_z(t) = \int\!\!\!\int\!\!\!\int_{V}
\left({n_i}_z (t) + \tilde{N_i}_{z}(t)\right)\, \mathrm{d} x\,
\mathrm{d} y\, \mathrm{d} z\ = \sum_i \left({n_i}_z (t) +
\tilde{N_i}_{z}(t)\right).
\end{equation}Again, the torque can be read as
the superposition of terms at dc, $1 f_m$ and $2 f_m$.

\subsection{Remnant moment~$\vec{m}_r$}\label{s:1m:m}
The test mass remnant magnetic moment $\vec{m}_r$ can be measured
from the torque it feels by an \emph{uniform} oscillating magnetic
field $\vec{B}_{0} \sin{\omega_m t}$. The torque in \eref
{e:mag:total:torque} becomes then
\begin{eqnarray}\label{e:mag:mom:torque}\nonumber
N_z(t) = \sum_i {n_i}_z (t) = \left(\sum_i d
\vec{m}_r(x_i,y_i,z_i)\right) \times \left(\vec{B}_{\textrm{res}}
+ \vec{B}_{0} \sin{\omega_m t}\right) = \\
= \vec{m}_r \times \left(\vec{B}_{\textrm{res}} + \vec{B}_{0}
\sin{\omega_m t}\right)
\end{eqnarray} and can be detected by coherent demodulation of the
fiber angular deflection $\phi(t)$ at the excitation frequency
$f_m = \omega_m / 2\pi$, through the torsion pendulum transfer
function
\begin{equation}\phi(\omega) = N(\omega) F(\omega) = N(\omega)
[\Gamma (1-(\omega/\omega_0)^2+ i/Q)]^{-1}.
\label{e:transfer:function}\end{equation} Here $\omega_0 =
\sqrt{\Gamma/I}$ is the pendulum resonance frequency, $\Gamma =
\Gamma_\textrm{f} \cdot (1+i \delta)$ is the torsional spring
constant, $Q = 1/\delta$ is the pendulum mechanical quality
factor, given by the inverse of the fiber loss angle $\delta$.
This experiment is routinely performed as a preliminary step to
evaluate the impact of the laboratory magnetic noise on high
sensitivity torsion pendula
performance~\cite{hueller:thesis,carbone:cqg}.
\begin{figure}[t]
\begin{center}
  \includegraphics[height=6cm]
  {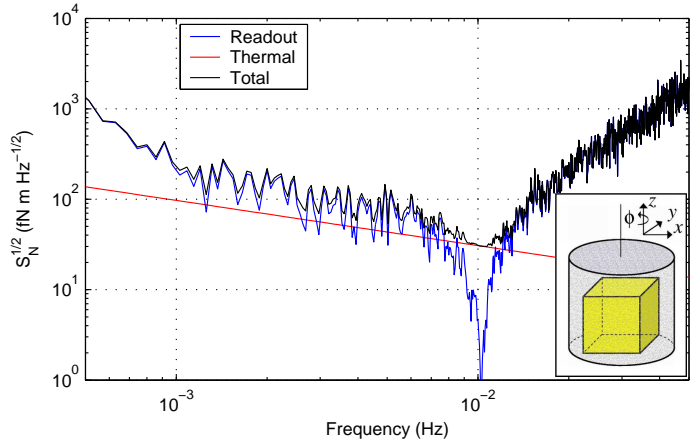}\\
  \caption[Single mass torsion pendulum: torque noise]{
  Expected torque noise floor for a torsion pendulum, comprised of
a commercial W fiber with length $L = 1$~m,
  diameter $d = 125$~$\mu$m, supporting along its axis one LISA
  cubic test mass, with side $L = 46$~mm, mass $M =
1.95$~kg, enclosed in a cylindrical Al holder
  as sketched in the inset.
  The pendulum thermal noise, conservatively assuming a quality factor
  $Q =1000$, is combined with
  the angular noise of the commercial autocollimator currently installed
  on the torsion pendulum facility~\cite{hueller:thesis,carbone:cqg},
  converted into equivalent mechanical noise through the torsion pendulum
  transfer function (\ref{e:transfer:function}). This noise performance
  can be applied to the measurement of both the test mass remnant moment
  $\vec{m}_r$ and susceptibility $\chi$, described in \sref{s:1m:m} and
  \ref{s:1m:chi}, respectively.
  }
  \label{f:mag:tor:pend:1}
\end{center}
\end{figure}

The expected torque sensitivity of the proposed torsion pendulum,
assuming a thermal torque noise spectrum for a quality factor $Q =
1000$, is shown in \fref{f:mag:tor:pend:1}. The maximum
sensitivity is set by the combination of the optical readout and
the pendulum thermal noise at $S^{1/2}_N \approx$~\rthz{30}{fN m}
around 10 mHz. Applying an oscillating magnetic field with
amplitude $\left \vert \vec{B}_0 \right \vert = $ 10
$\mu$T\footnote[3]{This excitation field configuration can be for
instance produced by means of a set of Helmholtz coils with radius
$R = 45$~cm, wound with $n=40$~turns and powered with currents
$I=100$~mA. The coil size is sufficient to ensure that the field
gradient over the extension of the test mass is negligible.}, and
assuming 3 hour integration time, this noise performance yields an
expected remnant moment resolution of $\left\vert \Delta \vec{m}_r
\right\vert \approx 40$~pA$\:$m$^2$, well below the LISA
requirements. Suitable flipping of the test mass within the holder
will allow measurement of all components of the magnetic moment
$\vec{m}_{r}$. To take advantage of this resolution, the remnant
moment of the sample holder, without test mass, should be measured
and subtracted (typical values for torsion pendula are $\left\vert
\vec{m}_r \right\vert \approx 1-10$~nA$\:$m$^2$
\cite{hueller:thesis,hoyle}).

\subsection{Susceptibility~$\chi$}\label{s:1m:chi}\begin{figure}[!t]
\begin{center}
  \includegraphics[height=6cm]{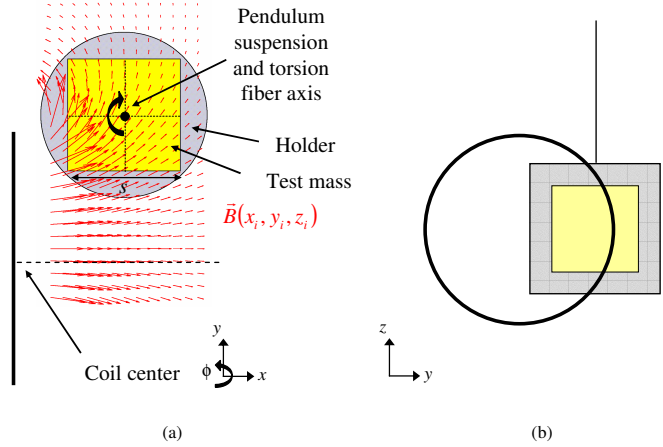}\\
  \caption[Single mass torsion pendulum for $\chi$ measurement]{
  Source coil arrangement for test mass susceptibility $\chi$ measurement. (a)
  Top view, with the test mass displaced with respect to the coil axis
  and to the coil plane. The red arrows represent the components of the
  applied magnetic field in the horizontal plane ${B_0}_x$ and ${B_0}_y$.
  (b) Side view. The single coil configuration creates a magnetic
  field and field gradient which differ on opposite sides of the test
  mass, inducing then a net $2 f_m$ torque ${N_{z}}_{2 f_m}(t) \propto
\chi$. The relative test mass - coil position has been chosen in
order to maximize the susceptibility induced torque (see
\fref{f:N:2w:y}).}
  \label{f:chi:scheme}
\end{center}
\end{figure}
\begin{figure}[t]
\begin{center}
  \includegraphics[height=6cm]
  {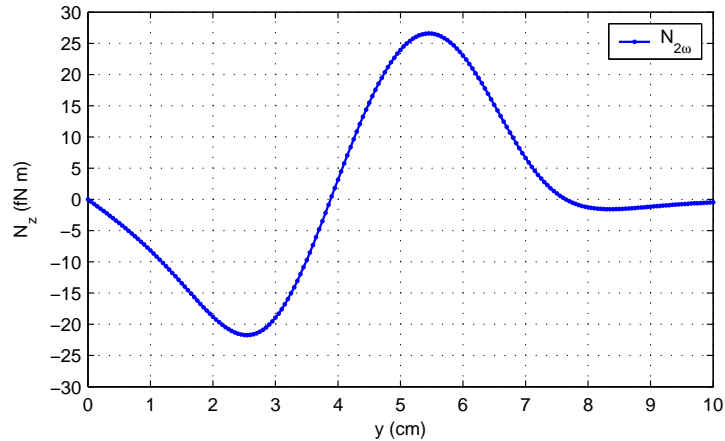}\\
  \caption[]{The susceptibility-induced signal
${N_{z}}_{2 f_m}(t)$ in \eref{e:mag:susc:torque}, due to the
interaction of the test mass with the magnetic
 field produced by the coil in the arrangement sketched in \fref{f:chi:scheme},
 as a function of the displacement $y$ between the test mass and the coil axis.
 The coil parameters are: radius $R = 5$~cm,
$n = 4$~turns, excitation current $I=2$~A, on-axis displacement $x
= 4$~cm; the maximum torque is obtained with off-axis displacement
$y = 5.5$~cm.}
  \label{f:N:2w:y}
\end{center}
\end{figure}

Assuming homogeneous test mass, the $2
f_m$torque component in \eref{e:mag:total:torque} can be written
as
\begin{equation}\label{e:mag:susc:torque} {N_{z}}_{2 \omega_m} =
-\frac{\chi \mathcal{V}}{2 \mu_0} \sum_i \vec{B}_{0}(\vec{x})
\cdot \left(x_i \frac{\partial}{\partial y} \vec{B}_{0}(\vec{x}) -
y_i \frac{\partial}{\partial x} \vec{B}_{0}(\vec{x})\right) \cos{2
\omega t}.
\end{equation}Applying a forcing magnetic field pattern which exerts
different forces on opposing test mass sides, it is then possible
to single out the effect induced by the susceptibility $\chi$ as a
net torque signal ${N_{z}}_{2 \omega_m}$ at twice the excitation
frequency $f_m$ directly proportional to the test mass average
susceptibility $\chi$. Any effect from the remnant moment will not
directly couple to this measurement and will appear only in
${N_{z}}_{1 \omega_m}$. A realistic configuration, compatible with
the dimensions of the facility vacuum vessel, is described in
\fref{f:chi:scheme}, and employs a relatively small coil, with
symmetry axis placed away from the pendulum torsion axis. In the
case of a test mass average susceptibility $\chi = 3\cdot
10^{-6}$, the expected induced torque at twice the excitation
frequency is shown in \fref{f:N:2w:y}, and has a maximum of
$\left\vert {N_{z}}_{2 \omega_m}\right\vert \approx 26.5$~fN\ m.
Assuming the same torque noise level and a 3 hour measurement as
in \sref{s:1m:m}, this signal amplitude will permit a resolution
$\left\vert \Delta \chi \right\vert \approx 5 \cdot 10^{-8}$,
corresponding to $\lesssim 2$\% of the LISA goal.

The measurement resolution grows as $\vec{B}_{0}
    \cdot \frac{\partial}{\partial x} \vec{B}_{0} \propto I^2$, so it is
possible to effectively improve the signal to noise ratio by
increasing the source coil drive current $I$. However, the major
uncertainty is given here by the model assumed to evaluate the
magnetic field and field gradients, and thus to estimate the
forces and torques. Nevertheless, the power of the technique is
evident because it will allow characterization of the LISA test
mass directly from the torques exerted by time-depending magnetic
fields, and independent estimate of the material proprieties to be
compared with other characterization methods.

\section{Conclusions}
The wide range of properties requested for the LISA test masses
(good optical quality to serve as end mirrors of the
interferometers, stringent machining tolerances to avoid stray
cross coupling, high mechanical strength to sustain the launch
vibrations, high density and homogeneous composition to minimize
acceleration for given force, good magnetic cleanliness) makes its
production a fundamental process within the LISA technology
development. In addition, the strict requirements on the magnetic
cleanliness make the verification of these proprieties a very
important issue. Even if a comprehensive testing campaign is
planned during the preliminary phases of the flight, with the aim
of establishing the ``feedthrough'' of the magnetic field
fluctuations to acceleration noise~\cite{LTP:2}, a ground based
experimental investigation of the force/torques associated with
varying magnetic field is highly desirable. The torsion pendulum
technique, with its widely demonstrated high torque sensitivity,
can be applied to a significant characterization of the magnetic
proprieties of the LISA/LTP test masses. The principle of
operation, based on the coherent detection of small torques
associated with modulation of external magnetic fields, is
analogous to the test procedure to be employed during the flight,
and thus represents an important validation step in view of the
mission. Work is currently in progress to modify the existing
facility in order to host the proposed experiment, and modelling
is being performed in order to assess the validity of the analysis
and the method performance.

\ack It is a pleasure to acknowledge many fruitful discussions
with E Adelberger, J Mester, D L Gill, S Anza, A Sanchez, D Chen.

\end{document}